\pdfoutput=1
\documentclass[sigconf, dvipsnames]{acmart}

\acmConference[ICSE 2024]{46th International Conference on Software Engineering}{April 2024}{Lisbon, Portugal}
\AtBeginDocument{%
  \providecommand\BibTeX{{%
    \normalfont B\kern-0.5em{\scshape i\kern-0.25em b}\kern-0.8em\TeX}}}

\acmConference[ICSE 2024]{46th International Conference on Software Engineering}{April 2024}{Lisbon, Portugal}
\usepackage{amsmath, amsfonts}
\usepackage[ruled,linesnumbered]{algorithm2e}
\usepackage{graphicx}
\usepackage{textcomp}
\usepackage[frozencache=true,cachedir=./minted-cache]{minted}
\usepackage{subcaption}
\usepackage{simplebnf}
\usepackage{pifont} 
\usepackage{xspace}
\usepackage{tcolorbox}     %%%%% colorbox
\usepackage{threeparttable}
\usepackage[export]{adjustbox}
\usepackage{multicol}
\usepackage{multirow}
\def\CVE{62 }
\def\TEST{3,631 }
\newcommand{\appname}{\textbf{\textit{PS}$^3$}\xspace}
\copyrightyear{2024} 
\acmYear{2024} 
\setcopyright{acmlicensed}\acmConference[ICSE '24]{2024 IEEE/ACM 46th International Conference on Software Engineering}{April 14--20, 2024}{Lisbon, Portugal}
\acmBooktitle{2024 IEEE/ACM 46th International Conference on Software Engineering (ICSE '24), April 14--20, 2024, Lisbon, Portugal}
\acmDOI{10.1145/3597503.3639134}
\acmISBN{979-8-4007-0217-4/24/04}
\begin{document}

%%
%% The "title" command has an optional parameter,
%% allowing the author to define a "short title" to be used in page headers.
\title[PS$^3$: Precise Patch Presence Test based on Semantic Symbolic Signature]{PS$^3$: Precise Patch Presence Test based on \\ Semantic Symbolic Signature}
%%
%% The "author" command and its associated commands are used to define
%% the authors and their affiliations.
%% Of note is the shared affiliation of the first two authors, and the
%% "authornote" and "authornotemark" commands
%% used to denote shared contribution to the research.

\author{Qi Zhan}
\orcid{0000-0002-6800-1857}
\affiliation{%
  \institution{The State Key Laboratory of Blockchain and Data Security, Zhejiang University}
   \city{Hangzhou}
  \country{China}
}
\email{qizhan@zju.edu.cn}

\author{Xing Hu}
\orcid{0000-0003-0093-3292}
\authornote{Corresponding Author}
\affiliation{
  \institution{The State Key Laboratory of Blockchain and Data Security, Zhejiang University}
   \city{Hangzhou}
  \country{China}
  }
\email{xinghu@zju.edu.cn}

\author{Zhiyang Li}
\orcid{0009-0000-5632-9479}
\affiliation{
  \institution{Zhejiang University}
   \city{Hangzhou}
  \country{China}
}
\email{misakalzy@zju.edu.cn}

\author{Xin Xia}
\orcid{0000-0002-6302-3256}
\affiliation{
  \institution{Software Engineering Application Technology Lab, 
  Huawei}
   \country{China}
  }
\email{xin.xia@acm.org}

\author{David Lo}
\orcid{0000-0002-4367-7201}
\affiliation{
  \institution{Singapore Management University}
  \country{Singapore}
}
\email{davidlo@smu.edu.sg}

\author{Shanping Li}
\orcid{0000-0003-2615-9792}
\affiliation{
  \institution{The State Key Laboratory of Blockchain and Data Security, Zhejiang University}
   \city{Hangzhou}
  \country{China}
}
\email{shan@zju.edu.cn}

%%
%% By default, the full list of authors will be used in the page
%% headers. Often, this list is too long, and will overlap
%% other information printed in the page headers. This command allows
%% the author to define a more concise list
%% of authors' names for this purpose.
% \renewcommand{\shortauthors}{Qi et al.}

%%
%% The abstract is a short summary of the work to be presented in the
%% article.
\begin{abstract}
During software development, vulnerabilities have posed a significant threat to users.
Patches are the most effective way to combat vulnerabilities. 
In a large-scale software system, testing the presence of a security patch in every affected binary is crucial to ensure system security.
Identifying whether a binary has been patched for a known vulnerability is challenging, as there may only be small differences between patched and vulnerable versions.
Existing approaches mainly focus on detecting patches that are compiled in the same compiler options. 
However, it is common for developers to compile programs with very different compiler options in different situations, which causes inaccuracy for existing methods.
In this paper, we propose a new approach named \textbf{\textit{PS}$^3$}, referring to \emph{precise patch presence} test based on \emph{semantic-level symbolic signature}. \textbf{\textit{PS}$^3$} exploits symbolic emulation to extract signatures that are stable under different compiler options. Then \textbf{\textit{PS}$^3$} can precisely test the presence of the patch by comparing the signatures between the reference and the target at semantic level.

To evaluate the effectiveness of our approach, we constructed a dataset consisting of 3,631 (CVE, binary) pairs of 62 recent CVEs in four C/C++ projects. The experimental results show that \textbf{\textit{PS}$^3$} achieves scores of 0.82, 0.97, and 0.89 in terms of precision, recall, and F1 score, respectively.
\textbf{\textit{PS}$^3$} outperforms the state-of-the-art baselines by improving 33\% in terms of F1 score and remains stable in different compiler options.

\end{abstract}

%%
%% The code below is generated by the tool at http://dl.acm.org/ccs.cfm.
%% Please copy and paste the code instead of the example below.
%%

% \begin{CCSXML}
% <ccs2012>
%  <concept>
%   <concept_id>10010520.10010553.10010562</concept_id>
%   <concept_desc>Computer systems organization~Embedded systems</concept_desc>
%   <concept_significance>500</concept_significance>
%  </concept>
%  <concept>
%   <concept_id>10010520.10010575.10010755</concept_id>
%   <concept_desc>Computer systems organization~Redundancy</concept_desc>
%   <concept_significance>300</concept_significance>
%  </concept>
%  <concept>
%   <concept_id>10010520.10010553.10010554</concept_id>
%   <concept_desc>Computer systems organization~Robotics</concept_desc>
%   <concept_significance>100</concept_significance>
%  </concept>
%  <concept>
%   <concept_id>10003033.10003083.10003095</concept_id>
%   <concept_desc>Networks~Network reliability</concept_desc>
%   <concept_significance>100</concept_significance>
%  </concept>
% </ccs2012>
% \end{CCSXML}

% \ccsdesc[500]{Computer systems organization~Embedded systems}
% \ccsdesc[300]{Computer systems organization~Redundancy}
% \ccsdesc{Computer systems organization~Robotics}
% \ccsdesc[100]{Networks~Network reliability}

%%
%% Keywords. The author(s) should pick words that accurately describe
%% the work being presented. Separate the keywords with commas.

%% A "teaser" image appears between the author and affiliation
%% information and the body of the document, and typically spans the
%% page.

% \received{20 February 2007}
% \received[revised]{12 March 2009}
% \received[accepted]{5 June 2009}

\maketitle

\section{Introduction}
Software vulnerabilities are security issues that can cause software systems to be attacked or abused by malicious users. 
Vulnerabilities may allow hackers or malware to invade the system and steal sensitive information, or allow attackers to remotely control the system and perform malicious operations~\cite{vulnerability}. 
The number of newly discovered vulnerabilities has increased rapidly in recent years, according to CVE data~\cite{CVE}.

Applying patches is one of the most effective ways to combat vulnerabilities and improve system security~\cite{OPUS}. Software developers continuously release patches to fix known vulnerabilities. In a large-scale software system, there may be thousands of binary files, which may be different versions and developed by various vendors or developers.
Therefore, it is essential to ensure that these binary files have been patched for the corresponding vulnerabilities. The process to ensure this is referred to ``patch presence test''.

The concept of the patch presence test was first introduced by Zhang et al.~\cite{zhang2018precise} in 2018 to distinguish it from conventional vulnerability searching for binaries or binary function matching.
We can formalize the concept of patch presence test by describing its input and output. The input consists of three parts: (1) information about a particular patch, usually a patch file that describes the code modification. (2) The source code or the \emph{reference} binaries of the project. The \emph{reference} binaries typically include a binary compiled from a vulnerable version of the project and another binary compiled from the patched version. (3) The \emph{target} binary to be tested.
The output of the patch presence test is a binary decision on whether the specific patch is present in the target binary or not.

Determining whether a given binary file has been patched for a corresponding vulnerability is challenging. The semantic gap between the source code and the binary makes it difficult to directly correlate modifications in the source code with changes in the assembly code~\cite{zhang2018precise}. Additionally, patches are usually small and subtle. Li et al.~\cite{empirical} show that up to 50\% of security patches only modify less than 7 lines of code. The traditional methods of matching binary functions are no longer effective in detecting the existence of patches due to the small differences present between the vulnerable and patched functions.

Many works have been proposed~\cite{zhang2018precise, sun2021osprey, jiang2020pdiff, xu2020patch, xie2023java} to automate patch presence test. 
These approaches can be divided into two groups: similarity-based and signature-based.
Similarity-based approaches~\cite{jiang2020pdiff, xu2020patch, xie2023java} extract information from \emph{reference} binaries, then measure similarity to the \emph{target} binary and decide the presence of patches based on similarity between two \emph{reference} to \emph{target}. Signature-based approaches~\cite{zhang2018precise, sun2021osprey} extract signatures from modification and attempt to verify their existence in the target. Both the two types of approach rely heavily on syntactic information.
However, binaries compiled with different options can be quite different at syntax level~\cite{chen2013influences}. 
\textbf{As a result, existing methods are only effective when reference binaries are compiled with the same compiler options as target binaries, making it difficult to test the binaries when the options are different.}

In addition, compiler flags or optimization levels cannot be easily extracted from a compiled binary~\cite{opt_is_useful}. Existing approaches usually resort to tools such as BinDiff~\cite{bindiff} to identify reference binaries most similar to the target binary~\cite{zhang2018precise} or compiling all binaries with the same compiler options in experiments. Generating the reference binary most similar to the target takes a long time. Jiang et al.~\cite{jiang2020pdiff} noted that the use of BinDiff~\cite{bindiff} took six minutes in an end-to-end test. \textbf{Efficiency is limited in large-scale software systems with thousands of binaries to test.}

To address these problems, we propose \appname, with the aim of testing the target binary precisely when the compiler options differ. The differences that represent the semantic information of the patch are stable in all downstream binaries, making them useful in the patch presence test. To capture semantic information in assembly code, we need to understand its context. To compare signatures precisely, we are required to check the equivalence at a semantic level rather than a syntax level. Following the intuition above, \appname extracts the semantic signatures based on symbolic emulation and matches them at a semantic level.
The main idea of our approach is two-fold: 

\ding{182} In signature extraction, we use symbolic emulation to emulate the entire affected function and collect side effects (e.g., store data in memory) in a symbolic environment. 
From function entry, we execute the code in a symbolic environment and collect effects. Then we obtain four types of signatures from the effects of affected functions.

\ding{183} In signature matching, we compare the signature extracted from the target binary with references. We utilize a theorem prover to prove the equivalence of two signatures that differ in syntax but are equal at semantic level. 
\appname is capable of precisely testing the target binary with respect to the compiler options by matching the semantic level signatures.

The binaries in the existing dataset are always compiled with the same compiler options~\cite{xu2020patch}, which cannot reflect the performance of the approaches when the compiler option differs. To fill this gap, we construct a dataset consisting of \TEST (CVE, binary) pairs on \CVE CVEs of four popular C/C++ projects. A test in our dataset is to check if a specific patch exists in a specific binary. We compile the binaries with combinations of different optimization levels and different compilers in the experimental phase.

Experimental results show that our approach outperforms state-of-the-art approaches by 0.22 in terms of the F1 score. We also evaluate the results of our approach compared to BinXray~\cite{xu2020patch} for every combination of compiler and optimization level.
The findings of our study establish the reliability and accuracy of our approach, indicating that \appname is capable of accurately capturing semantic signatures and successfully matching them with target binaries.

\textbf{Contributions}: In summary, the main contributions of this paper can be summarized as follows:
\begin{itemize}
    \item We propose \appname, a signature-matching-based approach, which can precisely and efficiently test the presence of patch in target binaries by symbolic emulation. 
    \item We construct a dataset comprising \CVE CVEs and corresponding binaries compiled with various compiler options from four popular C/C++ projects, resulting in a total of \TEST (CVE, binary) pairs. 
    \item We systematically evaluate our approach on the dataset, and the results demonstrate its effectiveness in patch presence test with high accuracy. 
\end{itemize}

The remainder of the paper is organized as follows. Section~\ref{sec:example} presents the motivating example of our study. Sections~\ref{sec:approach} describe the design and implementation of the framework. 
Section~\ref{sec:expset} and Section~\ref{sec:exprres} discuss the evaluation steps and results of \appname. 
In Section~\ref{sec:discussion}, we discuss the threats to validity of our approach and provide a case study of some representative patches.
Section~\ref{sec:related} summarizes the research related to the field. We conclude the paper and discuss the future work in Section~\ref{sec:conclusion}.

\section{Motivating Example}
\label{sec:example}
\begin{figure*}
\centerline{\includegraphics[width=0.8\linewidth, frame]{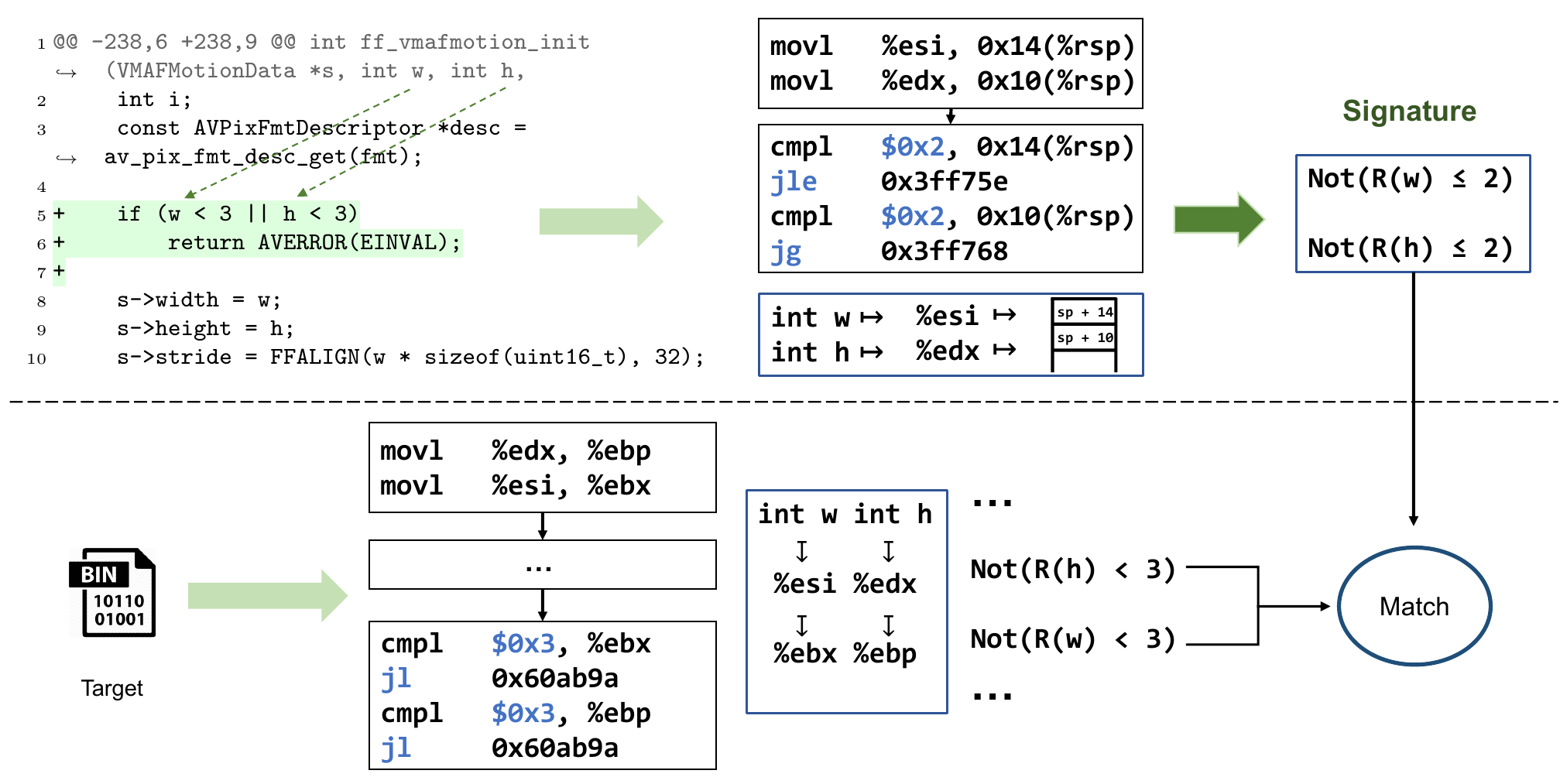}}
\caption{Motivating example}
\label{example}
\end{figure*}
In this section, we walk through a motivating example to illustrate our intuition. 
Figure~\ref{example} shows the testing process of the security patch for CVE-2020-22019~\cite{CVE-2020-22019}, a vulnerability in FFmpeg. The modification in the source code introduces a condition to check if the value in \verb|w| or \verb|v| is less than 3, where \verb|w| is the second parameter of the function \verb|ff_vmafmotion_init| and \verb|h| is the third parameter.

To test the presence of the patch, one straightforward idea is to find out what the binary code is compiled from the added condition and directly compare the extracted assembly code or similar pattern against the target binary.
The above line of source code might be compiled to \verb|cmpl $0x2, 0x14(%rsp); jle 0x1ab3b3|. However, matching the assembly code alone would not be helpful, as the concrete addresses and registers can vary across different binaries. For example, the target binary may use \verb|ebx| or \verb|ebp| as registers other than saving the value to stack. 
Additionally, the compare and jump pattern is very common in binaries, which can lead to many false positives if directly matched.

The above straightforward idea does not work well, as direct mapping of the source code to binary and matching cannot capture the semantic change of the patch.
This limitation arises from register allocation and the absence of contextual information. 
From the assembly code \verb|cmpl $0x2, 0x14(%rsp); jle 0x1ab3b3|, we know nothing about what the value pointed stands for in the current context. 
The requirement about context reminds us that we can track the whole dataflow in a symbolic environment from the function entry point.
To detect the status of the target binary after extracting the symbolic signatures, it suffices to test whether the signatures equivalent to \verb|w < 3| and \verb|h < 3| exist in the function.

Based on the idea mentioned above, we proceed with the following steps.

\textbf{Extraction.} 
We leverage symbolic emulation instead of concrete executing to track the changes in registers and memory relative to the function parameter and initial memory environment. After that, we extract signatures from the effects of the corresponding assembly code. In the prologue of the function shown in Figure~\ref{example}, the register represents the second parameter of the function saved to stack with offset \verb|0x14|. The code \verb|cmpl $0x2, 0x14(%rsp)|, which corresponds to condition \verb|w < 3| in the source code, can be traced back to the parameter \verb|w|. The code \verb|cmpl $0x2, 0x10(%rsp)| can be traced back to \verb|h| in the same way. As a result, we obtain the signatures \verb|Not(R(w) <= 2)| and \verb|Not(R(h) <= 2)|, where \verb|Not| denotes not jumping the branch.
The signatures remain constant in different binaries because the order of parameter passing is consistent with typical compiler behavior. Therefore, we can utilize them to precisely test the target binary.

\textbf{Matching.} 
Once the signature is generated, we perform the same symbolic emulation on the target binary function directly to collect effects. Since we cannot map the assembly code back to the source code in the target binary, all effects collecting in emulation are considered as potentially matched signatures. The remaining problem is how to match the effects collected from the target binary with the signatures. The signature \verb|Not(R(h) < 3)| does not appear to be exactly equivalent to \verb|Not(R(h) <= 2)| extracted from the target binary, but is semantically equal when \verb|R(h)| is an integer. With the use of a theorem prover, we are able to ensure that the two expressions are equal at the semantic level. By matching the two expressions, we decide that the target binary is patched.

\section{Proposed Approach}
\label{sec:approach}
\begin{figure*}
\centerline{\includegraphics[width=0.9\linewidth]{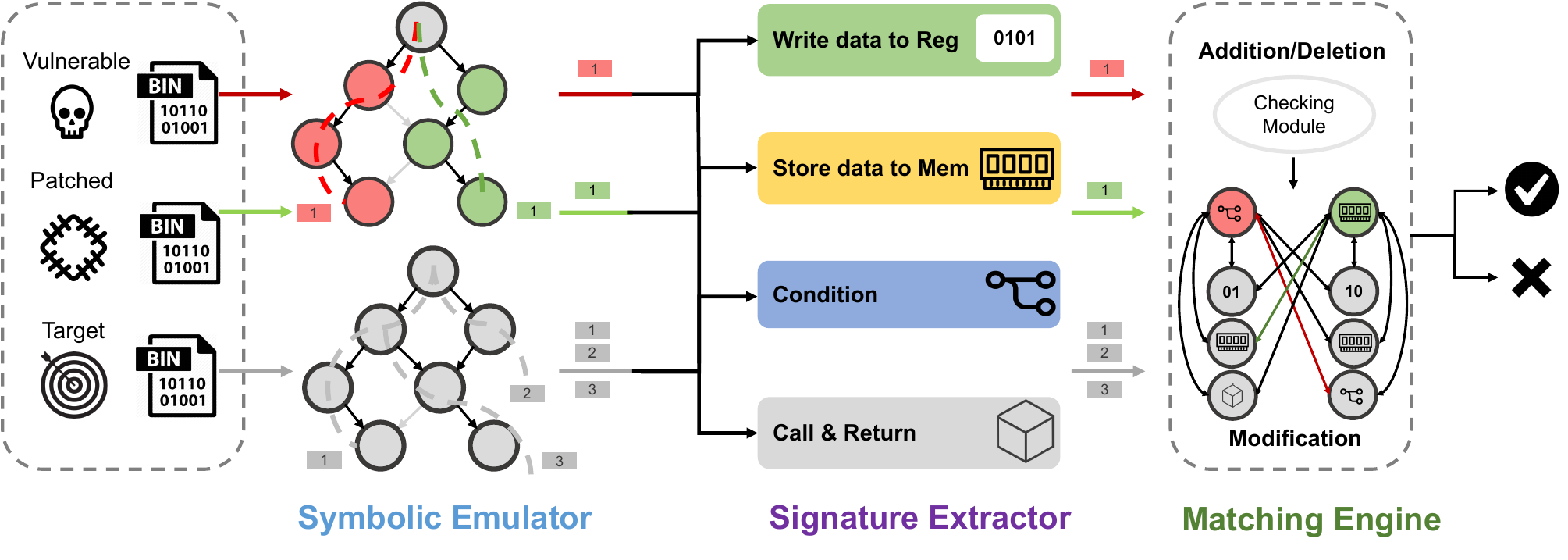}}
\caption{Overall framework of \appname}
\label{overall}
\end{figure*}
\subsection{Overall Framework}

The overall framework of \appname is illustrated in Figure~\ref{overall}. For simplicity, we assume that the patch modifies only one function. In practice, we classify the binary as vulnerable if one affected function is identified as vulnerable.
The main components of our approach include a symbolic emulator that emulates the function and generates side effects, a signature extractor responsible for extracting defined signatures from the effects, and a matching engine that determines the presence or absence of the patch by comparing the signatures.

\subsubsection{Symbolic Emulator} The emulator serves as the core of our approach. We first compile the source code into two reference binaries (that is, one vulnerable version and one patched version) with debug information and parse the difference file to extract the code modifications. Then, we utilize the debug information to identify which lines of reference binary code are deleted or added due to modifications in the source code. During the extraction phase, we perform symbolic emulation on the control flow graph of the binary function. For reference binaries, it is necessary to trace only the effects corresponding to the modified code, i.e., \textcolor{red}{red} curves for deleted code and \textcolor{Green}{green} curves for added code in Figure~\ref{overall}. For the target binary, we need to collect the \textit{complete} set of signatures in every possible affected function, i.e. \textcolor{gray}{gray} curves since we cannot map the binary code back to the source code by debug information.

\subsubsection{Signature Extractor} The extractor utilizes the emulator to perform symbolic emulation rather than concrete execution on the specific functions of the reference and target binaries, thus obtaining the effects. We extract these side effects as four types of signatures: writes data to register, stores data to memory, conditions, and calls or returns value. The formal definition of signatures is depicted below. Once the signatures have been extracted from the two reference binaries successfully, we can always use the same signatures for every target rather than searching for the most similar reference each time, saving a lot of time in practice.

\subsubsection{Matching Engine} In the matching phase, we compare the signatures of the target function with the two reference signatures one by one and determine whether the target binary has been patched or not.

\begin{algorithm}
\caption{Symbolic emulation procedure}
\label{algorithm}
\KwIn{Initial state, $s_i$}
\KwOut{Collected effects, $T$}  
\BlankLine
$Q \gets $ a queue of states initialized of $s_i$  \;
$visited \gets$ an array of booleans initialized to \textbf{false}\;
$T \gets$ $\emptyset$ \;
\While {$Q$ is not empty} { \label{dfs:begin}
    pop the last element $s$ from $Q$\;
    \If{$!visited[s]$}{ \label{dfs:visit}
        $visited[s]\gets$ \textbf{true}\;
        $next\_states$ = emulate\_basic\_block($s.block$)\;
        \For{$state\in next\_states$}{
            \eIf{$state$ is dead}{
                $T \gets T \bigcup state.trace$\; \label{dfs:trace}
            }{push $state$ to $Q$\;  \label{dfs:push}
            }
             
        }
    }
}\label{dfs:end}
\Return {$T$\;}
\BlankLine
% \SetKwProg{Pn}{Function}{:}{\KwRet state}  
  \SetKwFunction{FMain}{emulate\_basic\_block}
  \SetKwProg{Fn}{Function}{:}{}
  \Fn{\FMain{block}}{ \label{dfs:emulate}
        \While{state.successors less than 2} {
             \For{instruction in $block$}{
                state = emulate(instruction)\;
            }
        }
        \KwRet state\;
  }
\end{algorithm}

\subsection{Symbolic Emulation}
The Algorithm~\ref{algorithm} shows the high-level algorithm of symbolic emulation.
During the emulation phase, we start from the entry of the function and perform a deep first emulation (from line~\ref{dfs:begin} to line~\ref{dfs:end}). 
If a basic block is visited, it will be excluded during the next iteration (line~\ref{dfs:visit}). After emulating
a basic block, we push the successor states to the queue (line~\ref{dfs:push}) and collect side effects (line~\ref{dfs:trace}).

When emulating a basic block (line~\ref{dfs:emulate}), we emulate all the instructions in the basic block. All calls to other functions are skipped and a symbolic value is assigned to the register representing the return value, which records the function call name and parameters encountered during the emulation. We simply ignore all other possible side effects during this process.  
As mentioned before, the register and memory accessed can be represented by a symbolic value for the first time, e.g., function parameters accessing, and the actual operations on registers or memories are replaced by symbolic operations.

The choice of memory model has always been a problem in symbolic execution, since we cannot make a precise presumption for a symbolic pointer~\cite{symbolsurvey}. The problem still exists in symbolic emulation. We need to maintain a mapping between the symbolized addresses and their corresponding symbolic values, which ensures that we can correctly reuse the symbolic values when reading memory.
In \appname we use a simple memory model: Every symbolized address that is structurally different corresponds to a different symbolic value.
Considering the widely used stack to save local variables, the calculation involved with stack pointer is maintained separately with offset.

The main difference between symbolic emulation in our approach and traditional symbolic execution is that we do not aim to solve the constraints to obtain the concrete values that can reach specific basic blocks of the function. Our approach focuses on the semantic effects that a patch to the source code represents, rather than on constraint solving and concrete values, since the source code changes are always reachable. % However, the speed of symbolic execution can be impractical for real-world programs to reach the specific block affected by the patch commit.

\subsection{Signature Definition}

After generating the traces from the emulation, we need to extract useful information for matching. We define a binary signature as a group of side effects.
Inspired by Egele et al.~\cite{BLEX}, we choose the call to function, write to register or memory, condition statement, and return value as signatures.
We formalize the grammar of the signature in the Backus-Naur Form (BNF) as follows:

\begin{bnfgrammar}
  $\langle Signature  \rangle$ ::= $\langle FunctionCall  \rangle$ | $\langle RegisterWrite  \rangle$ | $\langle MemoryStore  \rangle$ | $\langle Condition  \rangle$ | $\langle Return  \rangle$
  ;;
  $\langle FunctionCall\rangle$ ::= name $\langle exp  \rangle$*
  ;;
  $\langle RegisterWrite  \rangle$ ::= offset $\langle exp  \rangle$
  ;;
  $\langle MemoryStore  \rangle$ ::=  $\langle exp  \rangle$ $\langle exp  \rangle$
  ;;
  $\langle Condition  \rangle$ ::= $\langle exp  \rangle$
  ;;
  $\langle Return  \rangle$ ::= name | index
  ;;
  $\langle exp  \rangle$ ::= $\langle exp  \rangle$ binop $\langle exp  \rangle$ | unop $\langle exp  \rangle$ | constant | symbol
\end{bnfgrammar}

$\langle FunctionCall\rangle$ is represented as the name of the function followed by its parameters, $\langle RegisterWrite\rangle$  is represented by the index followed by the value to be written, and $\langle MemoryStore  \rangle$ is represented by the address and the value to be written. A single Boolean expression represents a $\langle condition  \rangle$. If the name of the function call can be found in the symbol table, the returned value can be labeled accordingly. Otherwise, we assign an incremental index to it. Finally, an expression $\langle exp\rangle$ can be a combination of other expressions and operators, or it can be a constant value or a symbol mentioned in symbolic emulation.

The collection of write to register and store to memory is straightforward. We collect the condition information in every branch.
To collect information from the function call, we first obtain the name and number of parameters of a function call from the source code and use it to extract parameters from the actual function call. The return value is assigned after a function call.

There are many signatures to be extracted in the extraction phase as we record every possible behavior. However, some traces impact the decision, for example, common register write like ``\texttt{RegWrite R0, 1}''. Therefore, we implement certain rules to sanitize the signatures:
\begin{itemize}
    \item For the signatures of writing to memory and register, we remove the signature if the value has been used later to ensure our signatures are small and precise, which reduced the duplicate matches.
    
    \item In our approach, we prefer modified hunks over hunks that simply add or remove content, since there are unique signatures for the two references, respectively, and we can make a more precise comparison between the vulnerable and patched functions.  
\end{itemize}

\subsection{Signature Matching}

In the signature matching phase, we first check the extracted signatures before comparing them.
If the signatures in the patched binary contain signatures in the vulnerable binary, we call it\textit{ a pure addition}, since the change is essentially an addition of semantics. In the same way, we call it\textit{ a pure deletion} if the situation is the opposite.
If the modification is a pure addition or a pure deletion, we require all $\langle condition  \rangle$ and $\langle call  \rangle$ to match in the target binary.
The target is classified as vulnerable or patched if the requirement is not satisfied, respectively.
We do not decide the pure addition or deletion by looking at the difference file directly, since the result is easily affected by the syntax change in the source code. On the other hand, the semantic level signature is stable and they are not affected by syntax differences.

 After checking procedure, we compare the signatures extracted from the reference and target binary one by one to decide whether the target has been patched or not. The $match$ function shows the detailed comparison process. For the same type of signature, we compare every component of them and think that they are matched only when all components are equal. 
\begin{multline*}
\label{equal}
    match(s_1, s_2) = \\
    \begin{cases}
    \text{ } s_1.name = s_2.name \land \bigwedge_{i}s_1.para_i = s_2.para_i, & t_1=t_2=\text{ Call} \\ 
    \text{ } s_1.index = s_2.index \land s_1.value = s_2.value,  & t_1=t_2=\text{ RW} \\
    \text{ } s_1.addr = s_2.addr \land s_1.value = s_2.value, & t_1=t_2=\text{ MW} \\ 
    \text{ } s_1.exp = s_2.exp , &t_1=t_2=\text{ Condition} \\ 
    \text{ } s_1.name = s2.name \lor s_1.index = s_2.index, & t_1=t_2=\text{ Return} \\ 
    \text{ } \text{false}, & \text{otherwise} \\ 
    \end{cases}
    \\
    \text{where $t_1$ is the type of $s_1$, $t_2$ is the type of $s_2$.}
\end{multline*}

The score of reference to target is defined as the weighed sum of all matched signatures.
The matching of call expressions and condition expressions is much more important than writing to memory or registers, since they are more likely to be unique.
Therefore, we assign a higher weight to the matching of call and condition expressions than to the writes.
By comparing the weighted sum of matched signatures between the vulnerable and patched reference binaries, we decided whether the binary is patched. If the score of the vulnerable reference to the target is not less than patched, we classify the hunk as vulnerable. Considering the purpose of the patch presence test, we determine that the function is patched only if every hunk in the function has been patched and that binary is patched only if every affected function has been patched.

Furthermore, we also incorporate source code information into the signature matching.  
When the parameter of a function call is a string, we will ignore it in the signature match. 
Technically, we assign a wildcard symbol to the parameter, and all other values will be considered equal to it. For example, when matching the signature \verb|call(1, "string")|, the second parameter is ignored.
Additionally, when the patch code is contained in a conditional statement, which is a common pattern in vulnerability fixes, we deduce that the target binary must contain a matching signature for $\langle Condition \rangle$. If there is no matching signature of the condition type, we directly classify the target as vulnerable. The patch code in the motivating example is contained in \text{if (w<3 || h<3)}, so the target can be classified as vulnerable at once if there is no signature equal to \verb|w<3| or \verb|h<3|.

\subsection{Implementation}
\appname is built on top of the VEX intermediate representation (IR)~\cite{vexir} supported by angr~\cite{shoshitaishvili2016sok}. 
VEX IR is an intermediate representation of the instruction sets used in the Valgrind dynamic binary instrumentation framework~\cite{nethercote2007valgrind}. 
\appname can work with various architectures, since we can access the source code and compile it into the corresponding binaries.
Angr is an open source binary program analysis framework in Python and has been used extensively in binary analysis~\cite{symbion, eckert2018heaphopper, chen2020koobe}, as well as patch presence test~\cite{zhang2018precise, sun2021osprey, jiang2020pdiff}.
We use Z3~\cite{z3} to simplify the expression and calculation of the stack address offset to ensure that the memory mapping is more precise. In signature matching, we also use Z3 to prove equality of signatures nontrivial equation as in the motivating example.

\section{Experimental Setup}
\label{sec:expset}
We evaluate the effectiveness of our approach following four research questions.

\begin{table*}
\caption{Statistics of our dataset}
\label{stat}
\resizebox{\linewidth}{!}{%
\begin{tabular}{|c|c|c|cc|cc|cc|cc|cc|cc|cc|cc|c|}
\hline
\multirow{2}{*}{Projects} & \multirow{2}{*}{\#CVE} & \multirow{2}{*}{\#Version} & \multicolumn{2}{l|}{Clang \& O0} & \multicolumn{2}{l|}{Gcc \& O0} & \multicolumn{2}{l|}{Clang \& O1} & \multicolumn{2}{l|}{Gcc \& O1} & \multicolumn{2}{l|}{Clang \& O2} & \multicolumn{2}{l|}{Gcc \& O2} & \multicolumn{2}{l|}{Clang \& O3} & \multicolumn{2}{l|}{Gcc \& O3} & \multirow{ 2}{*}{\#Test} \\ \cline{4-19} 
                          &                        &                            & \multicolumn{1}{c|}{\#V} & \#P & \multicolumn{1}{c|}{\#V} & \#P & \multicolumn{1}{c|}{\#V} & \#P & \multicolumn{1}{c|}{\#V} & \#P & \multicolumn{1}{c|}{\#V} & \#P & \multicolumn{1}{c|}{\#V} & \#P & \multicolumn{1}{c|}{\#V} & \#P & \multicolumn{1}{c|}{\#V} & \#P &       \\ \hline \hline
OpenSSL             &        23                &     32                       & \multicolumn{1}{c|}{143}    &  161   & \multicolumn{1}{c|}{147}    &  163   & \multicolumn{1}{c|}{137}    & 135    & \multicolumn{1}{c|}{144}    &   152  & \multicolumn{1}{c|}{137}    &  135   & \multicolumn{1}{c|}{132}    & 114    & \multicolumn{1}{c|}{137}    &  150   & \multicolumn{1}{c|}{132}    &   114  &   2,233    \\ \hline
FFmpeg                    &        26                & 37                      & \multicolumn{1}{c|}{75}    &  115   & \multicolumn{1}{c|}{75}    &   115  & \multicolumn{1}{c|}{51}    &  76   & \multicolumn{1}{c|}{56}    &  78   & \multicolumn{1}{c|}{51}    &   76  & \multicolumn{1}{c|}{43}    &  71   & \multicolumn{1}{c|}{51}    & 76    & \multicolumn{1}{c|}{43}    & 71    &    1,123    \\ \hline
Tcpdump                   & 11                   & 4                       & \multicolumn{1}{c|}{33}    &   11  & \multicolumn{1}{c|}{33}    &  11   & \multicolumn{1}{c|}{21}    & 7    & \multicolumn{1}{c|}{24}    &  8   & \multicolumn{1}{c|}{21}    &   7  & \multicolumn{1}{c|}{24}    &  8   & \multicolumn{1}{c|}{21}    &  7   & \multicolumn{1}{c|}{21}    &  6   &  263      \\ \hline
LibXml2                   & 2                & 2                     & \multicolumn{1}{c|}{2}    & 0    & \multicolumn{1}{c|}{4}    &  0   & \multicolumn{1}{c|}{2}    &  0   &  \multicolumn{1}{c|}{1}    & 0    & \multicolumn{1}{c|}{1}    &   0  & \multicolumn{1}{c|}{0}    &   0  & \multicolumn{1}{c|}{1}    &  0   & \multicolumn{1}{c|}{1}    &   0  &   12    \\ \hline \hline
Total                     &      62                  &      75                      & \multicolumn{1}{c|}{253}    &  287   & \multicolumn{1}{c|}{259}    &  289   & \multicolumn{1}{c|}{211}    & 218    & \multicolumn{1}{c|}{225}    &  238   & \multicolumn{1}{c|}{210}    &  218   & \multicolumn{1}{c|}{199}    &  193   & \multicolumn{1}{c|}{210}    &   233  & \multicolumn{1}{c|}{197}    &  191   &   3,631     \\ \hline
\end{tabular}%
}
\begin{tablenotes}
\footnotesize
\item \#P represents the number of patched pairs while \#V represents vulnerable pairs. 
\end{tablenotes}

\end{table*}
\begin{itemize}
    \item \textbf{RQ1:} How effective is our approach at patch presence test compared to the state-of-the-art baselines?
    \item \textbf{RQ2:} How practical is our approach when testing binaries compiled with different options?
    \item \textbf{RQ3:} What is the result for each CVE and each project?

    \item \textbf{RQ4:} How efficient is \appname?
\end{itemize}

\subsection{Dataset}

\begin{figure}
\centerline{\includegraphics[width=\linewidth]{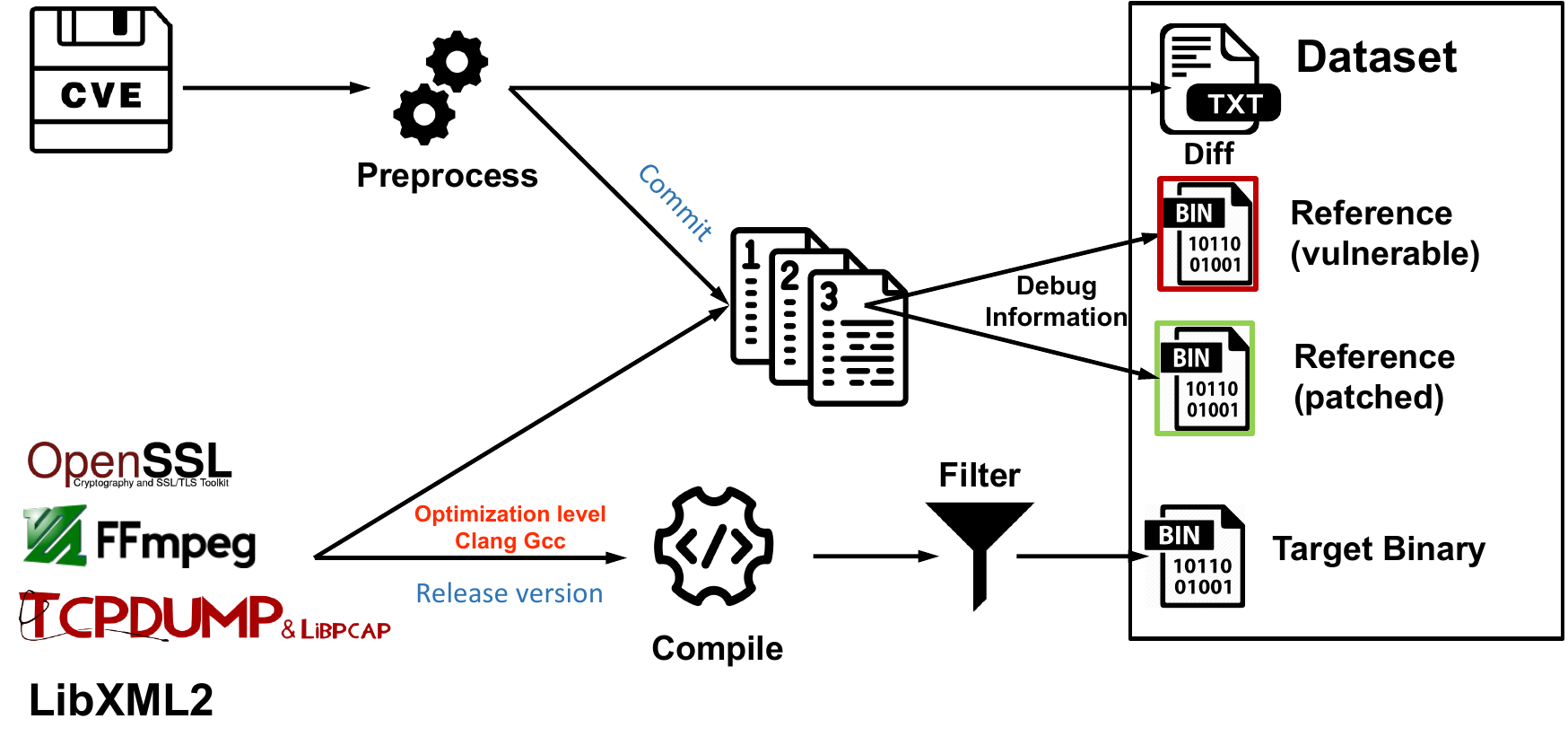}}
\caption{Dataset collection process}
\label{dataset}
\end{figure}

To evaluate our approach, we need to test binaries with various compiler options. The binaries in the existing dataset~\cite{xu2020patch, zhang2018precise} are always compiled with the same options. They cannot reflect the ability to test binaries with various compiler options. We collect CVE information to build our dataset, as depicted in Figure~\ref{dataset}.
We chose OpenSSL\footnote{https://www.openssl.org/news/vulnerabilities.html}, FFmpeg\footnote{https://ffmpeg.org/security.html}, Tcpdump\footnote{https://www.tcpdump.org/public-cve-list.txt}, and Libxml2\footnote{https://gitlab.gnome.org/GNOME/libxml2/-/releases} as our evaluation projects. 
The four projects involved cryptographic protocols, video processing, packet and xml analyzer. They are widely used in studies on vulnerability matching and patch presence testing~\cite{xu2020patch, sun2021osprey}. We only consider these four popular C/C++ projects in our experiments, although the method we proposed is general for compiled language.

For these projects, we extract vulnerability information from its official website instead of the CVE website, as it provides well-documented security updates. In OpenSSL, the affected version and the fix version are well documented. This allows us to directly determine the time of the first occurrence of a vulnerability and the corresponding patched version. 
For other projects, we manually search the code base and decide on the start scope affected by the vulnerability.
We selected CVEs from the past five years. For a vulnerability that exists in multiple branches, we choose only one branch for the test.

After collecting the CVE information and source code, we first generate difference files based on the patch commit. Then we compile the source code into two sets of binaries: reference binaries and target binaries.

To generate reference binaries, we compile the corresponding source code using gcc compiler with \texttt{O0} optimization level and \texttt{-g} for debug information. We compile two reference binaries for each CVE. One binary represents the patched version, which is compiled by the patch commit, and the other represents the vulnerable version, which is compiled by the commit just before the patch commit. 
    
To generate binaries for testing, we compile the corresponding source code using gcc and clang compilers with the optimization level ranging from \texttt{O0} to \texttt{O3} respectively, resulting in a total of \textbf{eight} different binaries.
In our experiments, the gcc version is 9.4.0 while the clang version is 13.0.0. Kim et al.~\cite{Kim_2023} have pointed out that the compiler version has a small impact compared to the optimization levels, so we only consider the optimization levels and ignore the different versions of the same compiler when generating target binaries.
We choose to compile the source code to every release version, instead of compiling from the patch commit, as we did for generating the reference binaries. In addition, we manually filter out binaries when affected functions are missing in symbol tables due to function inline.

We determine the ground truth using the affected version information extracted from CVE. We consider the version in the \textit{same branch} before the patched version as vulnerable and the patched and subsequent versions as patched. An item in our dataset is a (CVE, binary) pair that determine whether a specific binary contains a specific vulnerability, see Table~\ref{test example} for example.
\begin{table}
  \centering
  \caption{(CVE, binary) pair example}
  \label{test example}
  \resizebox{\linewidth}{!}{
  \begin{tabular}{|c|c|c|c|c|}
    \hline
    CVE & Target & Version & Option &Ground Truth \\ \hline
    2018-14468 & Tcpdump & 4.9.3 &O0 \& Clang   & patch \\ \hline
  \end{tabular}
  }
\end{table}
As a result, we collected a total of \CVE CVEs and created \TEST pairs, consisting of 1,582 vulnerable pairs and 1,893 patched pairs.  
Detailed statistical data on the created dataset for each combination of configurations for each project can be found in Table~\ref{stat}.

\subsection{Baselines}
To evaluate the effectiveness of our approach, we compare it with the following two baseline models:

\subsubsection{BinXray\cite{xu2020patch}} A state-of-the-art patch presence test tool without the presence of source code. BinXray uses basic block mapping to extract the signature of a patch by comparing vulnerable and patched binary programs. Then, it uses trace similarity to identify whether a target program is patched or not. 

\subsubsection{Asm2Vec\cite{asm2vec}} A deep learning-based binary clone search method that is designed to combat code obfuscation and compiler optimization. We chose it for comparison to demonstrate the effectiveness of directly utilizing binary code similarity for patch presence testing.

We choose not to select~\cite{jiang2020pdiff, sun2021osprey} as our baselines, since there are no open source implementations available, making it difficult to compare them with our approach. 
FIBER~\cite{zhang2018precise} is primarily designed for patch presence tests on Android kernel images in AArch64 architecture. Sun et al.~\cite{sun2021osprey} pointed out that the signature generated by FIBER is kernel related and not suitable for other binaries, which cannot be directly compared with our approach using our dataset. 
\subsection{Metrics}
Since the patch presence test is essentially a binary classification task, we select precision (P), recall (R) and F1 score (F1) as metrics to evaluate our approach with other baselines.  

\textbf{Precision} refers to the case where the test correctly classifies a vulnerable binary. We define $R_R$ as the number of truly vulnerable binary detected, $R_I$ as the number of patched binary recognized as vulnerable, then precision can be expressed as:
\begin{equation*}
    P = \frac{R_R}{R_R+R_I}
\end{equation*}

\textbf{Recall} is the ratio of the number of actual vulnerable binary detected to the number of all vulnerable binary.
$R_R$ is defined as the number of truly vulnerable binary retrieved, and $R_N$ is the number of all vulnerable binary in the dataset. Recall can be defined as follows:
\begin{equation*}
    R = \frac{R_R}{R_N}
\end{equation*}

\textbf{F1 Score} stands for the overall performance of the test, represented by the harmonic mean of precision and recall. Given P as the precision and R as the recall, the F1 score is calculated as follows: 
\begin{equation*}
    F1=\frac{2PR}{P+R}
\end{equation*}

\subsection{Experimental Setting}
\appname does not require additional configurations.
We use IDA Pro binary, a disassembler tool~\cite{ida} to dump the binary code of the functions for BinXray, which is the same as the BinXray used in their original experiments.
To the best of our knowledge, no official open source implementation of Asm2Vec is available, so we utilized an unofficial implementation\footnote{https://github.com/oalieno/asm2vec-pytorch} followed~\cite{wang2022jtrans} with default parameter settings.
The experimental setup consists of an Intel CPU operating at 2.90GHz and running on Linux.

\section{Experimental Results}
\label{sec:exprres}
\subsection{RQ1: Our approach vs. Baselines}

To evaluate the effectiveness of \appname, we compare our approach with two state-of-the-art approaches, including BinXray~\cite{xu2020patch} and Asm2Vec~\cite{asm2vec} in terms of precision, recall, and F1 score using our dataset.
As shown in Table~\ref{RQ1}, our approach outperforms all state-of-the-art baselines in terms of precision, recall, and F1 score for practical purposes, achieving 0.82, 0.97, and 0.89 respectively. 

Asm2Vec~\cite{asm2vec} is a state-of-the-art approach that learns assembly representations. We directly use the function-level representation to match the target function. Asm2Vec cannot distinguish vulnerable and patched functions, and precision and recall are both about 50\%. The reason may be the lack of capturing small changes between vulnerable and patched functions, as we have previously proposed.

BinXray~\cite{xu2020patch} is a state-of-the-art patch-based vulnerability matching tool and performs better than Asm2Vec, which achieves an F1 score of 0.67 in our dataset. For the tested binaries BinXray returns ``Unknown'', we make the most conservative assumption for it, i.e., considering all binaries as vulnerable. 

\def\fiberp{0.36}
\def\fiberr{0.57}

\def\asmp{0.60}
\def\asmr{0.50}
\def\asmf{0.65}

\begin{table}
    \centering
  \caption{Effectiveness of our approach vs. baselines}
    \begin{tabular}{cccc}
        \toprule
    Approach & Precision & Recall & F1\\ \midrule
    Asm2Vec & \asmp & \asmr & \asmf \\ 
    BinXray$^*$ & 0.51 & 0.96 & 0.67 \\ 
    \appname & \textbf{0.82} & \textbf{0.97} & \textbf{0.89}\\
\bottomrule
    \end{tabular}
    \begin{tablenotes}
    \footnotesize
    \item[*] * In our experiments, BinXray returns the result of ``unknown'' for most targets when the compiler option combination is not gcc \& O0.
    We consider all ``unknown'' binaries vulnerable for practical purposes and present the corresponding results.
\end{tablenotes}
\label{RQ1} 
\end{table}

\begin{center}
\begin{tcolorbox}[colback=gray!10, arc=2mm, auto outer arc, boxrule=1pt] \textbf{Answer to RQ1:} \appname can effectively identify the patch in the target functions with an F1 score 0.89. It outperforms the state-of-the-art approaches, Asm2Vec and BinXray, by 37\% and 33\%, respectively.
\end{tcolorbox}
\end{center}

\subsection{RQ2: Specific Compiler Optimizations and Compilers} 
To evaluate the effectiveness of our approach in different compiler options, we analyze the precision, recall and F1 score for every combination, as Table~\ref{RQ2} shows. We have the following findings:

    \textbf{Finding 1.} The precision and F1 score of our approach are always higher than the results of BinXray in each combination of compiler and optimization levels, showing that \appname can test the presence of patch. Both \appname and BinXray perform best on the combination of O0 \& gcc, it is understandable since the reference binary is compiled with O0 \& gcc.
    
    \textbf{Finding 2.}  The precision of our approach is affected by optimization levels, the F1 score falling within the range of 0.93 to 0.90 in the gcc compiler and 0.89 to 0.86 in the clang compiler, respectively. Although our result cannot ignore the differences between all optimization levels and always make the right decision, the situation migrated. The compiler also affected the accuracy of \appname, as different compilers generate and optimize assembly and can be quite different~\cite{Kim_2023}.
    
    \textbf{Finding 3.} When analyzing binaries compiled with different compiler options, BinXray drops to only 50\% precision, while it achieves 0.74 for O0 \& gcc. BinXray cannot determine whether the binary is vulnerable or patched. The results indicate that BinXray can accurately determine vulnerability or patch status only when the target and reference binary are compiled using the same options, while in other situations it is impractical.

Compared to BinXray, which can even decide the status of binaries with different compiler options most of the time, \appname can actually capture the key semantic differences of patches rather than syntax differences in binaries, leading to a more precise decision with different compiler options.

\begin{table*}
    \centering
    \caption{Results on different compiler options}
    \label{RQ2}
    \begin{tabular}{ccccccc}
    \toprule
    {\textbf{Compiler option combination}} & \multicolumn{3}{c}{BinXray$^*$}               & \multicolumn{3}{c}{\appname} \\

 & Precision & Recall & F1       &  Precision & Recall & F1 \\ \midrule
Gcc \& O0 &0.74 &0.99 &0.85                    & 0.91 & 0.94 & 0.93 \\
Gcc \& O1 & 0.47&0.97 &0.63                             &   0.83 & 1 & 0.91 \\
Gcc \& O2  &0.51 &0.95 &0.66                           & 0.82 & 1 & 0.90 \\
Gcc \& O3 & 0.51& 0.95 & 0.66                 & 0.82 & 1 & 0.90\\
Clang \& O0 &0.46 & 0.97 & 0.63            & 0.83 & 0.95 & 0.89\\
Clang \& O1 &0.49 &0.92 & 0.64              & 0.80 & 0.96 & 0.87\\
Clang \& O2 &0.48 & 0.96& 0.64                & 0.77 & 0.96 & 0.86\\
Clang \& O3 &0.47 & 0.96& 0.63                      & 0.77 & 0.96 & 0.86
\\  \bottomrule
    \end{tabular}
    \begin{tablenotes}
    \footnotesize
    \item[*] \quad\quad\quad\quad\quad\quad\quad\quad\quad\quad\quad\quad* We considering the ``unknown'' response as vulnerability in BinXray as we mentioned in RQ1.
\end{tablenotes}

\end{table*}
\begin{center}
\begin{tcolorbox}[colback=gray!10, arc=2mm, auto outer arc, boxrule=1pt] \textbf{Answer to RQ2:} \appname maintains high performance with different compiler options. Our approach yielded the greatest improvement in F1 score when compiled with gcc and O1 optimization level, resulting in a 44\% increase.
\end{tcolorbox}
\end{center}

\subsection{RQ3: Results on Different Projects and CVEs}
\begin{figure}
\centerline{\includegraphics[width=1.0\linewidth]{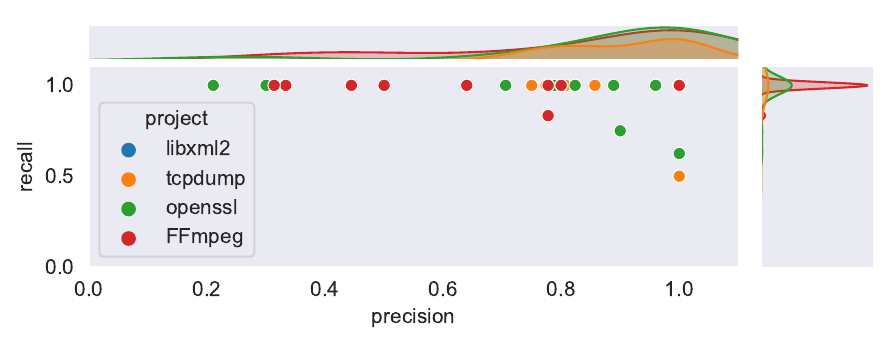}}

\caption{Results of every CVE in different projects}
\label{RQ3}
\end{figure}
To better understand the results of the experiments, we also list the precision and recall of each CVE used in our dataset in Figure~\ref{RQ3}. For 35 out of 62 CVEs, \appname accurately distinguishes between vulnerable and patched versions, which achieve a perfect F1 score. The recall of 58 CVEs is 100\% and the precison is lower respectively, since our approach is strict to decide a binary patched for practical use. In addition, our approach keep stable across various projects.
 
Our approach fails to identify four CVE (F1 score lower than 0.6). We check out each CVE manually and find that the five patches consist of one arithmetic optimization, one mentioned in the limitation below, and others are unclassified correctly since there is too much diff to confuse \appname.

\begin{center}
\begin{tcolorbox}[colback=gray!10, arc=2mm, auto outer arc, boxrule=1pt] \textbf{Answer to RQ3:}
\appname achieves a 100\% recall rate for 94\% of the CVEs and a 100\% F1 score for 56\% of the CVEs.
\end{tcolorbox}
\end{center}

\subsection{RQ4: Efficiency of our approach}

We evaluate the efficiency of \appname by measuring the time cost. 
Table~\ref{RQ4} shows the cost time of our approach and other baselines. 

\textbf{Preprocessing Time.} Preprocessing is not required in our approach since \appname takes binaries and patch file as input directly. On the other hand, the preprocessing time of IDA Pro required by BinXray for each file is 110 seconds in our experiments. It takes an average of 257 seconds to extract affected functions in a file in the FFmpeg project.

\textbf{Test Time.} It takes around 7 seconds to complete a test in our approach. The test with maximum time cost is about 45 seconds. We manually check the test. It is a vulnerability of function \verb|ff_mpeg4_decode_picture_header| in FFmpeg. There are a total of 570 basic blocks in the function, so emulation of the entire function can cost. In addition, signature matching is slower due to the increased number of signatures to be matched.
        Compared to our approach, BinXray is much faster in the test procedure, on average 60 milliseconds.

It seems that our approach is fairly slow compared to BinXray in terms of test time, but our approach is faster in practice considering the long preparation time in BinXray. In addition, a 7.4 seconds cost for determining a patch presence of a specific binary is acceptable for practical use. Furthermore, the signatures generated by our approach can be applied to multiple binaries directly, resulting in significant time savings.
\begin{center}
\begin{tcolorbox}[colback=gray!10, arc=2mm, auto outer arc, boxrule=1pt] \textbf{Answer to RQ4:} \appname can test the patch presence of binaries with an average of 7.4 seconds, without the time-consuming preprocessing.
\end{tcolorbox}
\end{center}
\begin{table}
  \caption{Time cost of our approach vs. BinXray (seconds)}
  \begin{center}
       \begin{tabular}{|c|c|c|c|c|}
       \hline
    Approach & Preprocess & Max & Min & Average \\
    \hline
    \appname & - &  45.1 &   2.1 & 7.4\\ \hline
    BinXray & 110 & 0.1 & - & 0.06 \\ \hline
\end{tabular}
\label{RQ4} 
  \end{center}

\end{table}

\section{Discussion}
\label{sec:discussion}
\subsection{Qualitative Analysis}
In this section, we show the strength and limitation of some representative security patches in our dataset.

\begin{figure}[htbp]
\begin{subfigure}{\columnwidth}
\centering
\begin{minted}[frame=single, fontsize=\scriptsize, breaklines, linenos, numbersep=2pt]{diff}
@@ -206,8 +206,8 @@ int ec_scalar_mul_ladder(const EC_GROUP *group, EC_POINT *r,
      */
     cardinality_bits = BN_num_bits(cardinality);
     group_top = bn_get_top(cardinality);
-    if ((bn_wexpand(k, group_top + 1) == NULL)
-        || (bn_wexpand(lambda, group_top + 1) == NULL)) {
+    if ((bn_wexpand(k, group_top + 2) == NULL)
+        || (bn_wexpand(lambda, group_top + 2) == NULL)) {
         ECerr(EC_F_EC_SCALAR_MUL_LADDER, ERR_R_BN_LIB);
         goto err;
     }
@@ -244,7 +244,7 @@ int ec_scalar_mul_ladder(const EC_GROUP *group, EC_POINT *r,
      * k := scalar + 2*cardinality
      */
     kbit = BN_is_bit_set(lambda, cardinality_bits);
-    BN_consttime_swap(kbit, k, lambda, group_top + 1);
+    BN_consttime_swap(kbit, k, lambda, group_top + 2);
 
     group_top = bn_get_top(group->field);
     if ((bn_wexpand(s->X, group_top) == NULL)
\end{minted}
\caption{Patch of CVE-2018-0735~\cite{CVE-2018-0735} }
\label{const}
\end{subfigure}

\vskip 0.5em

\begin{subfigure}{\columnwidth}
\centering
\begin{minted}[frame=single, fontsize=\scriptsize, breaklines, linenos, numbersep=2pt, breakanywhere]{diff}
@@ -100,6 +100,8 @@ void OPENSSL_LH_flush(OPENSSL_LHASH *lh)
         }
         lh->b[i] = NULL;
     }
+
+    lh->num_items = 0;
 }
\end{minted}
\caption{Patch of CVE-2022-1473~\cite{CVE-2022-1473} }
\label{bounded}
\end{subfigure}

\vskip 0.5em

\begin{subfigure}{\columnwidth}
\centering
\begin{minted}[frame=single, fontsize=\scriptsize, breaklines, linenos, numbersep=2pt]{diff}
@@ -59,9 +59,10 @@ static int ocsp_verify_signer(X509 *signer, int response,

     ret = X509_verify_cert(ctx);
     if (ret <= 0) {
-        ret = X509_STORE_CTX_get_error(ctx);
+        int err = X509_STORE_CTX_get_error(ctx);
+
         ERR_raise_data(ERR_LIB_OCSP, OCSP_R_CERTIFICATE_VERIFY_ERROR,
-                      "Verify error: %s", X509_verify_cert_error_string(ret));
+                      "Verify error: %s", X509_verify_cert_error_string(err));
         goto end;
     }
     if (chain != NULL)
\end{minted}
\caption{Patch of CVE-2022-1343~\cite{CVE-2022-1343} }
\label{limitation}
\end{subfigure}
\caption{Case study examples}
\end{figure}
\textbf{Constant change.}
It is common to change constant operands in a patch, see Figure~\ref{const} for example. It may be hard to identify this, as neither the control flow graph nor the data flow is altered. To address this situation, \appname can extract semantic changes to detect such differences. For CVE-2018-0735, \appname can discover that the parameter of the function call \verb|bn_wexpand| and \verb|BN_consttime_swap| change from \verb|group + 1| to \verb|group + 2|, then matches it with the signatures extracted from target binaries.

\textbf{Add bounded code.}
Adding code to ensure a certain bound is also a common pattern in vulnerability patches, see Figure~\ref{bounded} for example, which only requires one single line to fix the vulnerability. Similarly to our motivating example, \appname is capable of tracing symbolic transitions and extracting the semantic signature \verb|Mem(24 + SR(72)) == 0|, where \verb|SR(72)| corresponds to the first function parameter and \verb|24| corresponds to the corresponding structure offset in C. The target binary is tested by matching the single signature precisely.

\textbf{Limitation.}
\appname only considers forward instructions based on the control flow graph and terminates the emulation after traversing all relevant basic blocks in reference binaries.
In cases where the root of the vulnerability lies backward, \appname is unable to identify the differences and test the binary correctly.
For example, CVE-2022-1343~\cite{CVE-2022-1343}, a vulnerability in OpenSSL, is shown in Figure~\ref{limitation}. The vulnerability is a wrong assignment to \verb|ret|, which is later used as a return value. The difficulty in testing this patch is that one can only decide the existence of the patch after analyzing the backward instructions.
Since both the vulnerable and patched functions call the function \verb|X509_STORE_CTX_get_error| with the same argument, i.e. \verb|ctx|, \appname cannot distinguish between the two different functions. As a result, our approach extracts the same signature for both functions, resulting in labeling the binary as vulnerable even if it is patched.

\subsection{Threats to Validity}

\subsubsection{Internal Validity}

We used the \verb|CFGFast| method from angr~\cite{shoshitaishvili2016sok} in our implementation to build the control flow graph of a given function, which is fast but imprecise and may affect the results. Building a control flow graph for binaries is not easy and is sometimes difficult to analyze, in general~\cite{noteasy}. Additionally, due to the limitation of VEX IR, our approach may encounter additional problems in practice. Angr is not able to handle all vector instructions effectively, and OpenSSL heavily utilizes vector instructions to accelerate execution in the target binary. As a result, symbolic value transitions in emulation may not work correctly because symbolic register values cannot maintain valid information, leading to imprecise predictions.

In our approach, we assume that the entry address of the function to be analyzed is already known, which is also assumed in the related work~\cite{zhang2018precise, xu2020patch}. However, in practice, there are many stripped binaries. It is difficult to identify the entry point of a specific function in those binaries. To handle such cases, we can utilize binary function matching techniques as a preprocessing step in our approach to find the corresponding function, as previous work suggests~\cite{zhang2018precise, jiang2020pdiff}.

\subsubsection{External Validity}
We only consider vulnerabilities in four C/C++ projects, which may cause certain types of unusual vulnerability to be out of consideration.
As mentioned in \cite{v-szz}, the information about the affected version of the vulnerable software affected in the National Vulnerability Database (NVD)~\cite{CVE} is not always accurate, which can significantly affect the precision of our approach, although the information on the official project website is much more precise. In addition, \appname currently does not support an architecture other than Amd64, although it can be supported with some engineering effort.

\section{Related Work}
\label{sec:related}
\subsection{Patch Presence Test}

In this section, we review the main work closely related to patch presence test. The summary is shown in Table~\ref{related work}.
\begin{table}
    \centering
    \caption{Summary of related work}
    \label{related work}
    \begin{tabular}{lcccc}
        \toprule
    Work  &  \multicolumn{2}{c}{Approach} & Language & Source code \\  
        &    signature    & similarity &  & required \\ \midrule
        FIBER~\cite{zhang2018precise}  & \ding{51} &  & C/C++ & \ding{51} \\
        Osprey~\cite{sun2021osprey}  &  \ding{51} &  & C/C++ & \ding{51}\\
        PDiff~\cite{jiang2020pdiff}  &  & \ding{51} & C/C++ & \ding{51}\\
        BinXray~\cite{xu2020patch}  &  & \ding{51} & C/C++ & \\
        BSCOUT~\cite{dai2020bscout}  &  & \ding{51} & Java & \ding{51} \\
        PHunter~\cite{xie2023java}  &  &\ding{51} & Java &  \ding{51}\\
\bottomrule
    \end{tabular}

\end{table}

In 2018, Zhang et al.~\cite{zhang2018precise} first introduced the concept of ``patch presence test'' and distinguished it from conventional vulnerability search. Inspired by the heuristic that human analysts only examine the behavior of small and local code areas, Zhang et al. propose FIBER, which parses open source security vulnerability information to generate fine-grained binary signatures that reflect the changes caused by patch modifications. These signatures are used to determine whether the target binary has been patched. 
During the experimental phase, Zhang et al. evaluated FIBER using 107 real security patches from three different main vendors and eight images of the Android kernel, and the results showed that FIBER achieved an average accuracy of 94\% without false positives. 

Considering the overhead caused by the use of symbolic execution technology in FIBER~\cite{symbolic}, Sun et al.~\cite{sun2021osprey} proposed Osprey, which exploits light weight copy propagation and data flow slices without symbolic execution so that Osprey speeds up the testing process by more than 10 times without a much decrease in accuracy, which can still surpass 90\%.

Jiang et al.~\cite{jiang2020pdiff} proposed PDiff, a system that uses downstream kernel images to perform highly reliable patch presence testing. Unlike previous research on patch presence testing, PDiff is based on the semantic similarity of patch checking and thus has a high fault tolerance for code changes. Tests on 398 kernel images corresponding to 51 patches showed that PDiff can achieve high-precision testing with an extremely low miss rate.

We choose signature match instead of similarity-based, since our objective is to provide a definite answer. Compared to the above approaches for C/C++ projects, we use lightweight symbolic emulation instead of symbolic execution to avoid time-consuming constraint solving. Additionally, we are the first to utilize a theorem prover to check the equivalence of signatures at a semantic level, while other approaches do not.

The works mentioned above are based on the source code, while Xu et al.~\cite{xu2020patch} proposed BinXray, which does not presume the existence of the source code and patch commit. BinXray extracts patch signatures by comparing the given vulnerable code snippet and patched code snippet using a basic block mapping algorithm, where signatures are represented as a set of basic block traces. During the testing phase, the patch presence is determined by matching the basic block traces.

In addition to the detection of executable files compiled in the mainstream C/C++ languages mentioned above, Dai et al.~\cite{dai2020bscout} proposed BSCOUT, which is the first tool that can check the existence of patches in Java files. BSCOUT directly checks fine-grained patch semantics in the entire target executable without generating signatures. The results show that it exhibits significant accuracy regardless of the presence or absence of line number information (i.e., debug information) in the target executable. 

Xie et al.~\cite{xie2023java} proposed a tool named PHunter to address the challenge introduced by code obfuscation in Android applications. 
PHunter uses coarse-grained characteristics to identify functions related to patches and then evaluates semantic similarity to decide whether the code has been patched.

\subsection{Patch Presence Test vs. Other Works}

Although patch presence test, vulnerability detection~\cite{vulseeker, 8918305, 10.1145}, and binary function matching~\cite{BLEX,bingo, genius2016} share some similarities, they refer to different tasks in essential. We have summarized the differences between patch presence testing and vulnerability detection, binary function matching as follows:
\begin{enumerate}
    \item Compare with\textit{ vulnerability detection.} The essential difference between the patch presence test and conventional vulnerability detection is whether information about a particular patch is available or not. In vulnerability detection, one should decide whether the binary contains \textit{any} or a certain type of vulnerability. Patch presence test focuses on determining whether a \textit{specific} patch exists for a particular vulnerability.
    \item Compared with\textit{ binary function matching.} Patch presence test can be seen as a particular type of binary function matching method, that is, comparing the target function with the patched and vulnerable reference function and determining which one the target function is closer to. However, the general function matching methods do not work well due to the small differences between patched and vulnerable functions.
\end{enumerate}
\section{Conclusion and Future Work}
\label{sec:conclusion}
 Patch presence test is a critical task. Existing methods are heavily based on syntax information. We present \appname that utilizes symbolic emulation to generate semantic patch signatures for fast and precise matching.
To evaluate the performance of \appname, we created a dataset consisting of \TEST tests.
The experimental results show that our approach outperforms the baselines by improving 33\% in terms of F1 score. In addition, \appname performs well with different compiler options. 

When a patch is released to address a vulnerability in many Java projects, a corresponding test is often provided.
It reminds us to take advantage of dynamic execution-based testing methods to perform patch presence tests. %Patches that represent the change in semantic level can be reflected by specific inputs. 
Therefore, in the future, we plan to start with symbolic emulation and combine dynamic execution methods to construct concrete input to functions. %Execution can produce different side effects on vulnerable and patched functions in the same program environment. 
Running the function on the constructed input, we are able to precisely test the target binary.

The code and dataset we constructed are publicly available\footnote{https://github.com/Qi-Zhan/ps3}.

\section*{acknowledgments}

This research is supported by the Fundamental Research Funds for the Central Universities (No. 226-2022-00064), National Natural Science Foundation of China (No. 62141222), and the National Research Foundation, under its Investigatorship Grant (NRF-NRFI08-2022-0002). Any opinions, findings and conclusions or recommendations expressed in this material are those of the author(s) and do not reflect the views of National Research Foundation, Singapore.

\bibliography{ref}
\bibliographystyle{ACM-Reference-Format}

\end{document}